\def \SAIT #1 #2 {{\em Mem.\ Soc.\ Astron.\ It.\/} {\bf #1}, #2}
\def \MESS #1 #2 {{\em The Messenger\/} {\bf #1}, #2}
\def \ASTRNACH #1 #2 {{\em Astron. Nach.\/} {\bf #1}, #2}
\def \AAP #1 #2 {{\em Astron. Astrophys.\/} {\bf #1}, #2}
\def \AAL #1 #2 {{\em Astron. Astrophys. Lett.\/} {\bf #1}, L#2}
\def \AAR #1 #2 {{\em Astron. Astrophys. Rev.\/} {\bf #1}, #2}
\def \AAS #1 #2 {{\em Astron. Astrophys. Suppl. Ser.\/} {\bf #1}, #2}
\def \AJ #1 #2 {{\em Astron. J.\/} {\bf #1}, #2}
\def \ANNREV #1 #2 {{\em Ann. Rev. Astron. Astrophys.\/} {\bf #1}, #2}
\def \APJ #1 #2 {{\em Astrophys. J.\/} {\bf #1}, #2}
\def \APJL #1 #2 {{\em Astrophys. J. Lett.\/} {\bf #1}, L#2}
\def \APJS #1 #2 {{\em Astrophys. J. Suppl.\/} {\bf #1}, #2}
\def \APSS #1 #2 {{\em Astrophys. Space Sci.\/} {\bf #1}, #2}
\def \ASR #1 #2 {{\em Adv. Space Res.\/} {\bf #1}, #2}
\def \BAIC #1 #2 {{\em Bull. Astron. Inst. Czechosl.\/} {\bf #1}, #2}
\def \JSQRT #1 #2 {{\em J. Quant. Spectrosc. Radiat. Transfer\/} {\bf #1}, #2}
\def \MN #1 #2 {{\em Mon. Not. R. Astr. Soc.\/} {\bf #1}, #2}
\def \MEM #1 #2 {{\em Mem. R. Astr. Soc.\/} {\bf #1}, #2}
\def \PLR #1 #2 {{\em Phys. Lett. Rev.\/} {\bf #1}, #2}
\def \PASJ #1 #2 {{\em Publ. Astron. Soc. Japan\/} {\bf #1}, #2}
\def \PASP #1 #2 {{\em Publ. Astr. Soc. Pacific\/} {\bf #1}, #2}
\def \NAT #1 #2 {{\em Nature\/} {\bf #1}, #2}

\documentstyle{memsait}
\input epsf.sty
\begin{opening}
\title{X-ray Transients Observed with the RXTE All Sky Monitor 1996-1997}
\author{R. A. Remillard$^1$}
\institute{$^1$Massachusetts Institute of Technology, Cambridge, MA, USA}
\date{} 
\end{opening}

\begin{document}

\oddpagefooter{}{}{} 
\evenpagefooter{}{}{} 
\ 
\bigskip

\begin{abstract}
Highlights from the RXTE All Sky Monitor (ASM) during 1996 and 1997
are reviewed with particular attention to X-ray transients. The ASM
has detected 117 sources. These include 12 recurrent transients and 10
new X-ray sources, some of which began their outbursts before the
launch of RXTE. The majority of the outburst profiles are strikingly
different from the classical form, with its fast rise and slow
decay. Some of the light curves appear quasi-persistent, which
suggests that they are associated with secular changes in the
accretion rate from the donor star rather than with a cyclic
instability in the accretion disk. The nature of the compact object is
uncertain in many cases, but there is a likelihood that the majority
of new sources are black hole systems, while the majority of recurrent
transients contain neutron stars. There is a correlation
between the gross shape of the energy spectrum, as seen in the ASM
hardness ratio $HR2$, and the accretion subclass. These data, with
important contributions from SAX, CGRO, and Granat, provide a deeper
probe of the X-ray sky than has been available previously.  However,
there are important systematic effects to be investigated before a
meaningful evaluation of transient occurrence rates can be derived.

\end{abstract}

\section{Introduction}

The All Sky Monitor (ASM) on the {\it Rossi X-ray Timing Explorer}
(RXTE) has been regularly observing bright celestial X-ray sources
since 1996 February 22. Detector problems had been encountered during
the first days of operation (1996 January 5-12), but the instrument
has remained stable under an observing plan restricted to
low-background regions of the RXTE orbit (580 km altitude).  The ASM
currently operates with 20 of the original 24 detector anodes, and the
typical observation duty cycle is 40\%, with the remainder of the time
lost to the high-background regions of the orbit, spacecraft slews,
and instrument rotation or rewinds. The net yield from the ASM
exposures is about 5 celestial scans per day, excluding regions near
the Sun. The ASM instrument calibration, construction of the data
archive, the derivation of source intensities, and efforts to locate
new X-ray sources are all carried out with integrated efforts of the
PI team at M.I.T. and the RXTE Science Operations Center at Goddard
Space Flight Center.

The ASM instrument consists of three scanning shadow cameras (SSC)
attached to a rotating pedestal. Each camera contains a
position-sensitive proportional counter, mounted below a wide-field
collimator that restricts the field of view (FOV) to 6$^{o}$ x
90$^{o}$ FWHM and 12$^{o}$ x 110$^{o}$ FWZI. One camera (SSC3) points
in the same direction as the ASM rotation axis. The other two SSCs are
pointed perpendicular to SSC3. The latter cameras point toward a
common direction, but the long axes of their collimators are tilted by
+12$^{o}$ and -12$^{o}$, respectively, relative to the ASM rotation
axis. The ASM can be rotated so that the co-pointing SSCs are aligned
with the larger instruments of RXTE (i.e. the PCA and HEXTE). The top
of each collimator is covered with an aluminum plate perforated by 6
parallel (and different) series of narrow, rectangular slits.  These
slits function as a coded mask by casting a two-dimensional shadow
pattern onto the position-sensitive anode wires in the detector. The
histograms of accumulated counts from the anodes represent the
superposition of shadows from each X-ray source in the collimator's
field of view. Further information on this instrument is given by
Levine et al. (1996).

The ASM raw data includes 3 types of data products that are tabulated
and formatted for telemetry by the two ASM event analyzers in the RXTE
Experiment Data System. In the current observing mode, position
histograms are accumulated for 90 s ``dwells'' in which the cameras'
FOVs are fixed on the sky. Each dwell is followed by a 6$^{o}$
instrument rotation to observe the adjacent patch of sky. The rotation
plans for ASM dwell sequences are chosen to avoid having any portions
of the Earth in the FOVs of SSCs 1 and 2. The position histograms are
accumulated in three energy channels: 1.5--3.0, 3--5, and 5--12
keV. The second ASM data product consists of various measurements from
each camera binned in time. These data are useful in studying bright
X-ray pulsars, bursters, $\gamma$-ray bursts, and several other
categories of rapid variability. The ``good events'' from each camera
are recorded for each energy channel in 0.125 s bins, while 6
different types of background measures are recorded in 1 s bins.
Finally, 64-channel X-ray spectra from each camera are output every 64
s. These data provide a means of monitoring the detector gain, since
we may integrate the spectra over long time scales to observe the 5.9
keV emission line from the weak $^{55}$Fe calibration sources mounted
in each collimator. In addition, ASM spectra may be useful in
investigations of spectral changes in very bright X-ray sources.

The ASM data archive is a public resource available for both planning
purposes and scientific analysis. The source histories are available
in FITS format from the RXTE guest observer facility at
http://heasarc.gsfc.nasa.gov/docs/xte/xte\_1st.html. The ASM archive
is also available in ASCII table format from the RXTE web site at
M.I.T., http://space.mit.edu/XTE/XTE.html.

The majority of scientific applications for ASM data can be organized
into four categories: 
\begin{itemize}
\item Locating and Monitoring X-ray Transients,
\item Measuring Intensity and Spectral Variations in Persistent X-ray Binaries,
\item Long Term Behavior of Bright AGNs,
\item Positions and X-ray Properties of $\gamma$-Ray Bursts.
\end{itemize}

This paper briefly reviews the ASM results in the first category, with
consideration of results obtained during 1996 and 1997.

\section{X-ray Transients 1996-1997}

There has been a remarkable diversity of X-ray transients during 1996
and 1997. Table 1 summarizes 22 cases detected with the RXTE ASM;
these include both new sources and recurrent transients. We have
selected these to emphasize intrinsic (rather than geometric)
variations in accretion, and so we have excluded transients associated
with active coronae and X-ray sources with periodic (or nearly
periodic) recurrence intervals, such as systems with B-e type donor
stars in eccentric orbits.  Table 1 lists the known transients
observed with a peak X-ray flux above 20 mCrab (2--12 keV) during 1996
and 1997. This group includes recent discoveries from RXTE, SAX, CGRO,
and Granat. One of the strengths of the ASM program is the frequent
all-sky coverage and the relative ease with which historical light
curves may be extacted from the survey database in response to
discoveries from various researchers in the astronomical community.

\vspace{1cm} 
\centerline{\bf Table 1 - ASM Observations of X-ray Transients 1996-1997}

\begin{table}[h]
\hspace{1.5cm} 
\begin{tabular}{|l|c|c|c|c|c|c|c|}
\hline
\multicolumn{8}{|c|}{New and Continuing Transients} \\
\hline
\hline
Source    & type  & profile &peak & HR2  & start & $\Delta$T & comments \\
\hline
XTE J1716-389    & ns? &   qp &   66 & 0.9 & pre XTE &  -- & ASM detection\\
GRS 1737-31      & bhc &   qp &   26 & 1.6 & pre XTE &  -- & Cui 97\\
GRS 1739-278     & bhc & frsd &  805 & 0.6 &   1/96  & 270 & radio source \\
XTE J1739-302    & bhc &   qp &   34 & 1.3 & pre XTE &  -- & Smith 97 \\
SAX J1750-29     & ns  & frsd &  117 & 1.3 & 3/10/97 &  33 & qp?, Bazzano \\
XTE J1755-324    & bhc & frsd &  188 & 0.3 & 7/24/97 & 105 & Remillard 97\\
SAX J1808-3658   & ns  & frsd &  108 & 0.9 & 9/07/96 &  19 & in t'Zand 98\\
XTE J1842-042    & ns? &   qp &   21 & 1.2 & pre XTE &  -- & PCA detection\\
XTE J1856+053    & bhc &  sym &   75 & 0.4 & 4/16/96 &  27 & Marshall 97a\\
	''       &     & frsd &   79 & 0.4 & 9/02/96 &  70 & 2nd outburst \\
GRS 1915+105     & bhc &  irr & 2497 & 1.3 &   5/92  &  -- & huge var. \\
\hline
\\
\\
\hline
\multicolumn{8}{|c|}{Recurrent Transients} \\
\hline
Source    & type  & profile &peak & HR2  & start & $\Delta$T & comments \\
\hline
4U1210-64        & ns? &   qp &   30 & 1.2 & pre XTE &  -- & \\
X1354-644        & bhc &  sym &   52 & 1.3 &10/23/97 & $>$85 & \\
X1630-472        & bhc &  irr &  336 & 1.1 & 3/11/96 & 150 &  flat top \\
GRO J1655-40     & bh  &  irr & 3138 & 0.6 & 4/25/96 & 484 &  var. spectra \\
RX J17095-266    & ns? & frsd &  210 & 0.9 &12/31/96 &  86 & Marshall 97b\\
KS1731-260       & ns  &   qp &  356 & 1.1 & pre XTE &  -- & msec pulsar \\
Rapid Burster    & ns  & frsd &  377 & 1.5 & multiple &  25 & 3 outbursts \\
GRO J1744-28     & ns  & frsd & 1291 & 2.5 &12/95 12/96 &$>$120 & complex \\
GX 1826-238      & ns  &   qp &  120 & 1.2 & pre XTE &  -- & Bazzano 98 \\
X1845-024        & ns  &  sym &   24 & 2.6 & 9/96  5/97 &  21 & 2 outbursts\\
EXO 1846-031     & bhc &  sym &   30 & 1.5 &12/19/97 &   ? & \\
AQL X-1          & ns  &  sym &  370 & 0.9 & multiple & ~60 & 4 outbursts \\
\hline
\end{tabular}
\end{table}

In col. 4 of Table 1 we list the peak X-ray flux compiled using daily
ASM averages (mCrab at 2--12 keV), and in col. 3 the profile of the
outburst is described as either quasi-persistent (qp), irregular
(irr), symmetric (sym), or the more common shape with a fast rise and
slow decay (frsd). The starting date of the outburst is given (or
limited) in col. 6, while col. 7 lists the duration (days) for which
the X-ray emission is above the ASM threshhold, which is near 10 mCrab
at $3\sigma$ per one day time bin.

As indicated in col. 2, the X-ray transients of 1996-1997 include 1
confirmed black hole system (bh), 9 black hole candidates (bhc), 8
accreting neutron stars (ns), and 4 suspected ones (ns?).  All of the
confirmed neutron stars are bursters except for the pulsars GRO
J1744-28 and X1845-024. There are interesting correlations between the
X-ray spectrum and the type of accreting system, despite the coarse
energy resolution available from the ASM and the randomizing effects
of substantial differences in the amount of interstellar column
density among these sources. The ASM hardness ratio, $HR2$, is defined
as the flux in the 5--12 keV band relative to the flux in the 3--5 keV
band, and the mean value for each source is given in col. 5. Both
pulsars (with $HR2 \sim 2.5$) are significantly 'harder' than the
other sources.  The confirmed neutron star systems lie in the range
$0.9 \le HR2 \le 1.5$, while the black hole candidates have a bimodal
distribution, $HR2 \le 0.6$ or $HR2 \ge 1.1$. Since the accreting
black hole systems are known to exhibit composite spectra consisting
of a soft, thermal component and a hard, power-law component
(e.g. Tanaka \& Lewin 1995), the bimodal distribution in $HR2$ values
for the bhc can be understood as the optional dominance of the soft or
hard spectral component, respectively.

\section{Light Curves of Selected X-ray Transients}

In Figure 1 we show the ASM light curves for four examples among the
group of new X-ray transients. The first source was located with the
ASM during March of 1997.  The detection was gained from ASM sky maps
that are constructed on a weekly basis from the residuals of our data
analysis effort. For each camera - dwell, we fit the ASM position
histograms for the superposition of mask shadows of sources in the
camera's FOV, and then we compute the residuals from this fit. The
residuals histograms are then back-projected onto position cells fixed
on the celestial sphere. In each sky cell, the X-ray flux is computed
by cross correlating the residuals histograms against the modeled
shadow pattern for that particular cell during a given observation. In
the case of XTE J1716-389, the superposition of such residual flux
(while the source was unknown) yielded a significant detection at a
position (J2000) 259.10$^o$, -38.90$^o$, with an uncertainty radius
(90\% confidence) of 0.12$^o$. We then re-analyzed the ASM database
with routine inclusion of this new source position and extracted the
light curve shown in the top panel of Figure 1. This source is one of
7 quasi-persistent transients listed in Table 1. These cases exhibit
long-term (i.e. years) secular changes in the accretion rate, and they
appear very different from the transients associated with the
classical disk instability mechanism, which produces a wave of
accretion that generally causes an X-ray outburst of 1-6 months
duration (e.g. Cannizzo et al. 1995; Chen et al. 1997).

\begin{figure}
\epsfysize=16cm 
\hspace{0.5cm}\epsfbox{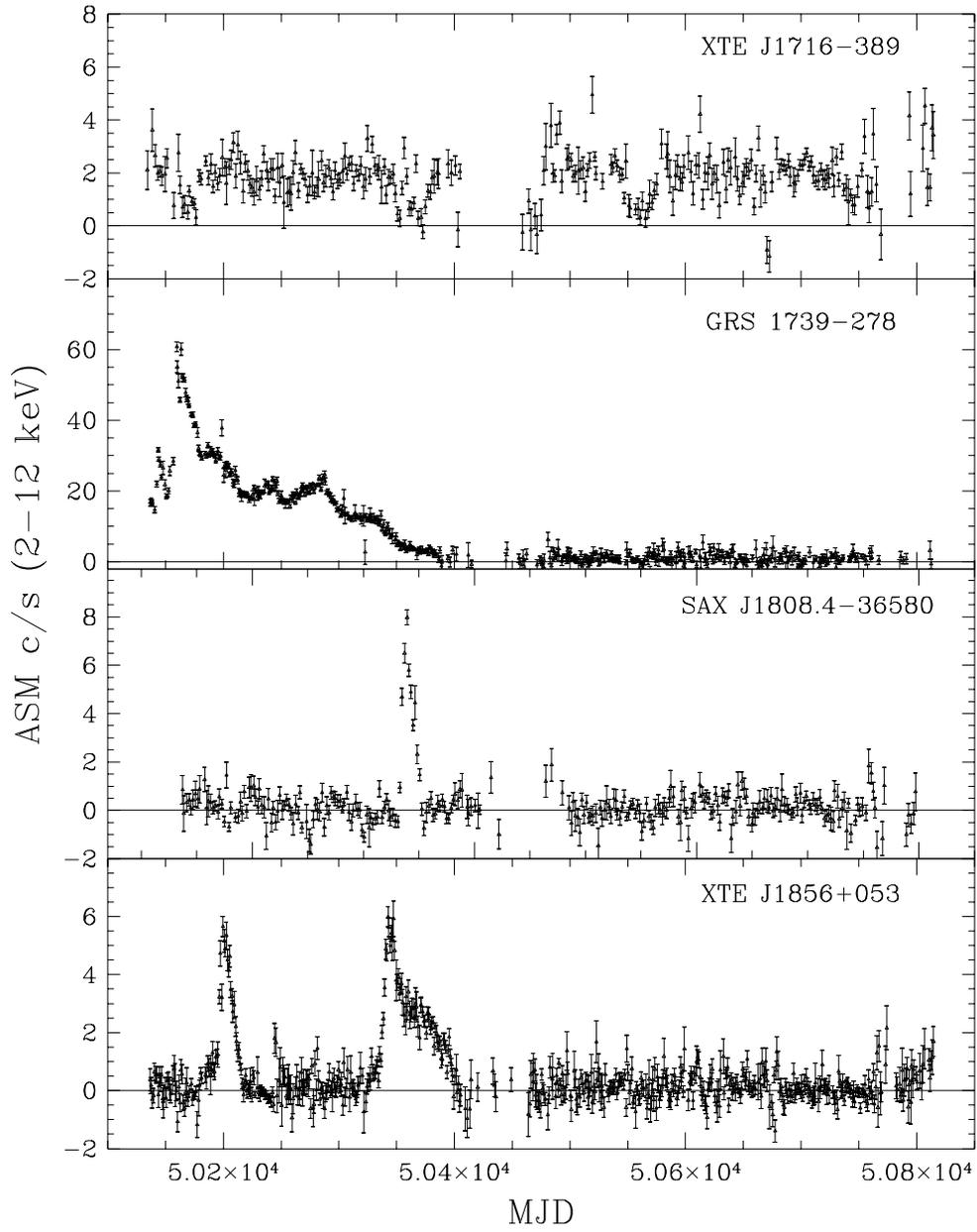} 
\caption[h]{ASM light curves for 4 new X-ray transients (1996-97). In the 2--12 keV band, the count rate for the Crab Nebula is 75.5 ASM c/s}
\end{figure}


The moderately bright outburst in GRS1739-278 (panel 2 of Figure 1)
displays the more typical 'frsd' outburst profile noted
above. However, in this case there are at least six decay phases after
X-ray maximum, and the sequence of accretion enhancements has extended
the outburst interval to $\sim$ 9 months. In contrast, the outburst
duration of SAX J1808.4-3658 (panel 3) is only 19 days. This X-ray
burster (in t'Zand et al. 1998) is one of several fast X-ray
transients that would likely have gone unnoticed without the current
complement of X-ray instruments that monitor the high-energy sky.
Finally, a source discovered originally in PCA scans of the galactic
plane, XTE J1856+053 (Marshall et al. 1997a) shows diverse decay
profiles in two X-ray maxima separated by only 4.5 months. Since this
interval is short relative to the subsequent span of X-ray quiescence,
it is possible that the two X-ray maxima are best regarded as
interrelated portions of a single X-ray nova episode.

Four light curves of recurrent transients are shown in Figure 2. The
bhc X1630-472 (top panel) exhibits a peculiar 'flat top' profile in
which the spectrum slowly hardens. The maximum luminosity occurs
$\sim$110 days after the initial X-ray rise, and it is not at all
clear how this may be explained employing the accretion disk
instability model. Even more peculiar is the light curve of the
microquasar, GROJ1655-40. It is considered here as recurrent transient
because of the extended period of X-ray quiescence during late 1995
and early 1996. A double wave in X-ray brightness is shown in panel 2
of Figure 2.  The strong flares in the first wave are absent in the
second wave, and the flares are entirely due to activity in the
power-law spectral component. We note that the gap in ASM coverage
between these waves, which is an annual feature also seen in other ASM
light curves, is caused by the Sun's passing near a given X-ray
source. The microquasars with relativistic radio jets, GROJ1655-40 and
GRS1915+105 (see Morgan et al. 1997) display the most complex light
curves in the ASM archive.

Panel 3 of Figure 2 shows the outburst in RX J1709.5-266 (Marshall et
al. 1997b). Very little had been known about this source apart from a
brief detection in the ROSAT survey. The ASM light curve suggests that
it is a disk-instability type of X-ray source, and it is a suspected
neutron-star system given the shape of its X-ray spectrum. Finally,
the most regularly recurring X-ray transient is Aql X-1 (see van
Paradijs 1995). Portions of 4 X-ray outbursts have been seen with the
ASM, and the average recurrence interval is $\sim 190$ days. The
second detection interval (MJD 50240--50310) appears to be some type
of failed outburst that presents yet another challenge
for the disk instability model.

The ASM detection threshhold for new X-ray transients ($\sim 25$ mCrab
in weekly residual maps) is higher than that for recurrent transients
($\sim 10$ mCrab per daily average), since the sky mapping technique
is substanitally less sensitive that the method for determining the
X-ray flux for known sources with accurate positions. Since many of
the new transients (e.g. the SAX discoveries) have brief outburst
duration, the systematic issues pertaining to completeness must be
investigated thoroughly in order to use the ASM archive to determine
reliable production rates for X-ray transients in the
Galaxy. Accordingly we plan to revise our instrument models and
conduct archival sky mapping efforts with a goal to understand the
completeness level for ASM detections of new sources, as a function of
both X-ray brightness and outburst duration.

\begin{figure}
\epsfysize=16cm 
\hspace{0.5cm}\epsfbox{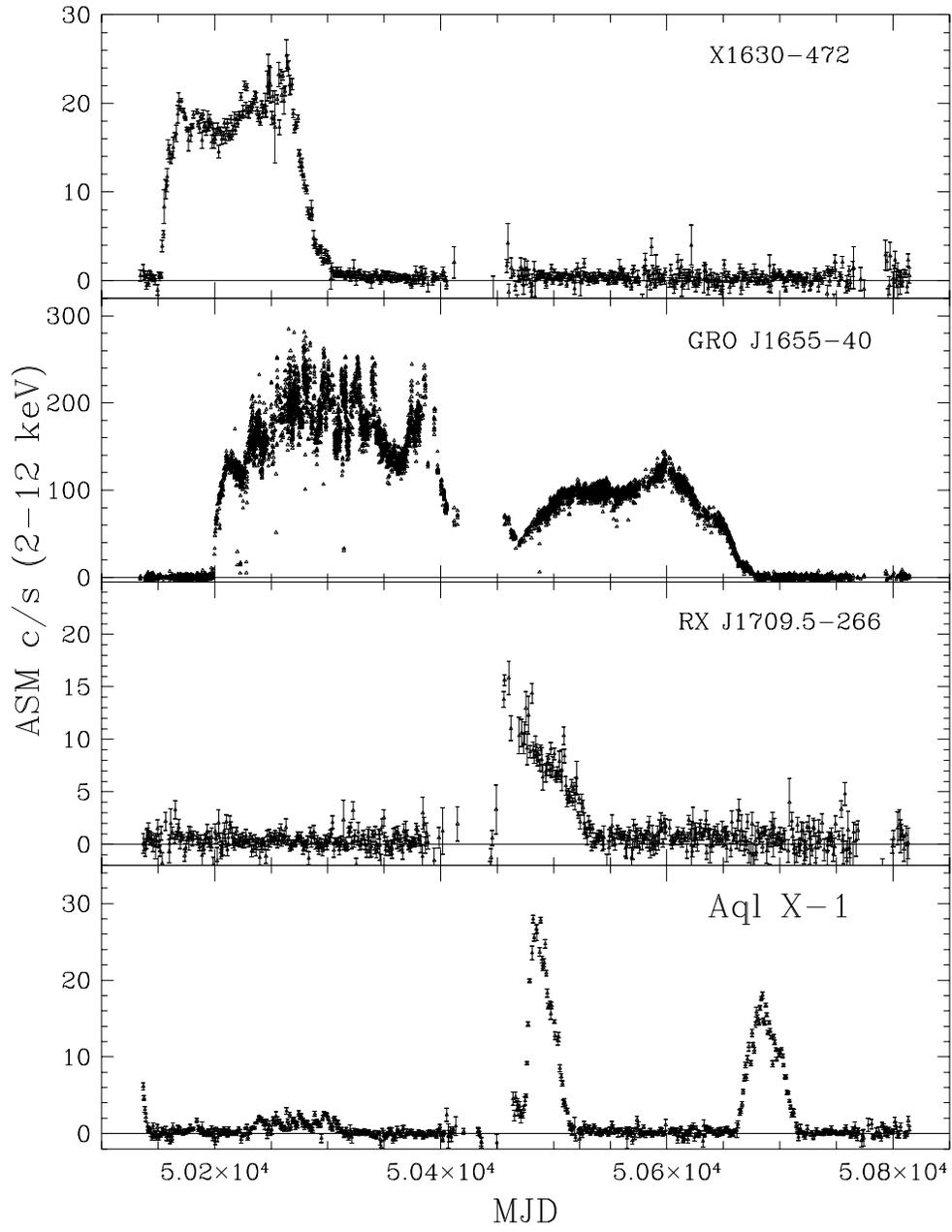} 
\caption[h]{ASM light curves (1996-97) for 4 recurrent transients.}
\end{figure}

\end{document}